\documentclass{article}

\usepackage{arxiv}
\usepackage[utf8]{inputenc} 
\usepackage[T1]{fontenc}    
\usepackage{hyperref}       
\usepackage{url}            
\usepackage{booktabs}       
 \usepackage{amsfonts}       
\usepackage{nicefrac}       
\usepackage{microtype}      
\usepackage{lipsum}

\usepackage{afterpage} 
\usepackage{etex}
\usepackage{color}
\usepackage[stable]{footmisc} 
\usepackage[section]{placeins} 
							
\usepackage{graphicx} 
\usepackage{multirow} 
\usepackage{longtable}

\usepackage{amsmath}
\usepackage{amssymb}
\usepackage{amstext}
\usepackage{amsthm}
\usepackage{enumerate}
\usepackage{extarrows}
\usepackage{esvect}

\usepackage{relsize}
\usepackage{array}
\usepackage{dsfont}

\usepackage{pstricks-add}
\usepackage{caption}
\usepackage{subcaption}	
\usepackage{float}

\usepackage{parskip}
\usepackage{ stmaryrd }
\usepackage{MnSymbol}

\usepackage[xcolor=dvipdf]{changes}

\usepackage{cancel}
\usepackage{framed} 
\usepackage{fancyhdr}
\usepackage{hyperref}

\newtheorem*{rem*}{Remark}

\newcommand{\N}{\mathbb{N}} 
\newcommand{\R}{\mathbb{R}} 
\newcommand{\C}{\mathbb{C}} 
\newcommand{\Id}{\mathcal{I}} 
\newcommand{\G}{\Gamma} %
\newcommand{\g}{\gamma} %
\newcommand{\e}{\text{e}} %

\newcommand{\B}{\mathcal{B}} %
\newcommand{\s}{\mathcal{S}} %
\newcommand{\E}{\mathcal{E}} %
\newcommand{\bp}{\mathbf{p}}

\newcommand{\bu}{\mathbf{u}}
\newcommand{\bw}{\mathbf{w}}
\newcommand{\bq}{\mathbf{q}}
\newcommand{\bv}{\mathbf{v}}
\newcommand{\zero}{\mathbf{0}}
\newcommand{\Kern}{\text{kern }} %
\newcommand{\Span}{\text{span }} %
\newcommand{\Order}{\mathcal{O }} %

\newcount\colveccount
\newcommand*\colvec[1]{
        \global\colveccount#1
        \begin{pmatrix}
        \colvecnext
}
\def\colvecnext#1{
        #1
        \global\advance\colveccount-1
        \ifnum\colveccount>0
                \\
                \expandafter\colvecnext
        \else
                \end{pmatrix}
        \fi
}

\definecolor{MyYellow}{rgb}{1.0, 0.75, 0.0}
\definecolor{MyGreen}{rgb}{0.0, 0.5, 0.0}
\definecolor{CrimsonGlory}{rgb}{0.75, 0.0, 0.2}
\definecolor{DarkRed}{rgb}{0.55, 0.0, 0.0}
\definecolor{ElectricPurple}{rgb}{0.75, 0.0, 1.0}

\definecolor{TUD_1a}{RGB}{93, 133, 195} 
\definecolor{TUD_1b}{RGB}{0, 90, 169} 
\definecolor{TUD_1c}{RGB}{0, 78, 138} 
\definecolor{TUD_1d}{RGB}{36, 53, 114} 

\definecolor{TUD_2a}{RGB}{0, 156, 218}
\definecolor{TUD_2b}{RGB}{0, 131, 204}
\definecolor{TUD_2c}{RGB}{0, 104, 157}
\definecolor{TUD_2d}{RGB}{0, 78, 115}

\definecolor{TUD_9a}{RGB}{233, 80, 62} 
\definecolor{TUD_9b}{RGB}{230, 0, 26} 
\definecolor{TUD_9c}{RGB}{185, 15, 34} 

\definecolor{TUD_3a}{RGB}{80, 182, 140} 

\definecolor{MyGreen}{rgb}{0.0, 0.5, 0.0}
\definecolor{MyPurple}{RGB}{141, 0, 255} 
\definecolor{MyCyan}{rgb}{0, 0.67, 0.67} 

\definecolor{TUD_11a}{RGB}{128, 69, 151} 

\definecolor{TUD_11b}{RGB}{114, 16, 133} 
\definecolor{TUD_11c}{RGB}{97, 28, 115}  
\definecolor{TUD_11d}{RGB}{76, 34, 106}  

\definecolor{TUD_10a}{RGB}{201, 48, 142} 
\definecolor{TUD_10b}{RGB}{166, 0, 132} 
\definecolor{TUD_10c}{RGB}{149, 17, 105}  
\definecolor{TUD_10d}{RGB}{115, 32, 84}  

\begin{document}

\title{ Obtaining the long-term behavior of Master equations with finite state space from the structure of the associated state transition network}

\author{
  Bernd Fernengel  \\
  Institut für Festkörperphysik \\
  Technische Universität Darmstadt \\
  Hochschulst. 6, 64289 Darmstadt, Germany \\
  \texttt{bernd@pkm.tu-darmstadt.de} \\
   \And
  Barbara Drossel  \\
  Institut für Festkörperphysik \\
  Technische Universität Darmstadt \\
  Hochschulst. 6, 64289 Darmstadt, Germany \\
  \texttt{drossel@pkm.tu-darmstadt.de} \\
}
\maketitle

\begin{abstract}
The Master equation describes the time evolution of the probabilities of a system with a discrete state space. This time evolution approaches for long times a stationary state that will in general depend on the initial probability distribution. 
Conditions under which the stationary state is unique are usually given as remarks appended to more comprehensive theories in the mathematical literature.  
We provide a direct and complete derivation of a necessary and sufficient criterion for when this steady state is unique. 
We  translate this problem into the language of graph theory and show that there is a one-to-one correspondence between minimal absorbing sets within the state-transition network and linearly independent stationary states of the Master equation. 
\end{abstract}


\section{Introduction}
Master equations appear in many different context in chemistry, biology and physics. Examples are biochemical reaction networks, stochastic population dynamics based on deaths and births, and various statistical physics problems. 

Master equations describe the time evolution of the probabilities of the different states of a system, with the dynamics being due to transitions between these states \cite{van1992stochastic,honerkamp2012statistical}. The most common form is an initial value problem of a linear differential equation. 
The picture behind a Master equation is that of probabilities flowing between states like a fluid, where their total amount is being conserved. The probability flow along a directed link depends linearly on the strength of the link and the amount of probability at the source.  
Of particular interest is the long-time behaviour of the solution, as this determines the states where a system will eventually be, together with their probabilities. Unlike discrete-time  Markov chains, which can lead to long-term oscillations of the probabilities, the solution of the Master equation for a finite state space will always converge to a  stationary solution where the probability distribution does not change any more. In general, this steady state will depend on the initial distribution of the probabilities, i.e.,  on the initial condition of the corresponding differential equation.

In this paper we focus on the question under which conditions a finite-size Master equation has a unique steady state, in which case the Master equation is called  \textit{relaxing} \cite{spohn1977algebraic}.

The mathematical literature contains necessary and sufficient conditions for the stationary state to be unique, but these are only hinted at or given as a remark appended to theorems within major treatises on topics such as Master equations, Markov chains or directed graphs  \cite{bremaud2013markov, privault2013understanding, douc2018markov}. 
What these all lack is a clear statement and proof that is based on first principles. 

In the physical literature, usually more special situations are discussed. For instance, the Master equation for the microcanonical ensemble of statistical physics can be shown to be relaxing because the transition rates between pairs of states are symmetric. Therefore, the steady state is the uniform distribution \cite{van1992stochastic,honerkamp2012statistical}. %

The often-cited review by Schnakenberg on the network theory of Master equation systems \cite{schnakenberg1976network} discusses models where to each transition there exists also the reverse transition. In this specific case, the stationary solution is always unique. 
A more mathematical paper by Jamilowski and Staszewski \cite{jamiolkowski1992master} gives a general criterion for the Master equation to be relaxing, however, this criterion is a nonintuitive mathematical expression based on the principal minors of the matrix that contains the transition rates.

Our goal in this paper is to provide a direct proof that is accessible for physicists without delving deep into the mathematical literature. 
This proof proceeds by translating the initial value problem into the language of graph theory. By considering the network of the states of the system and the transitions between them, an intuitive understanding of the meaning of the steps of the proof can be achieved. An important notion is that of \textit{minimal absorbing sets}, which are subsets of states from where the probability cannot escape. We show that the number of linearly independent steady states of the Master equation equals the number of minimal absorbing sets in the corresponding transition network. In particular, the steady state of the Master equation is unique if and only if there is exactly one minimal absorbing set.
While this statement seems intuitively clear, the complete proof is non-trivial, even though it uses only undergraduate mathematics.


\section{The Master equation and the time evolution operator}
We consider a system of $N \in \N$ states and the transitions between them. If there is no transition from state $i$ to state $j$, the associated transition rate $ \g_{i\to j}$ vanishes. The states and the nonzero transition rates form together a network $\s=(\Omega, \E)$, with the states $\Omega = \{1, \dotsc, N\}$ being the nodes and the transitions being the directed links $\E \subseteq \Omega\times \Omega$ of the network. The transition rates indicate the weights of these links. 
For all subsets of states $B\subseteq \Omega$ there is a corresponding subnetwork $(B, \E_B)$, where $\E_B := \{(i,j) \in \E \,:\, i,j \in B\}$ consists of all links  in the subset $B \subseteq \Omega$. 

Since transitions occur only between different states, the network has no self-loops, that is $\g_{i \to i} = 0$ for all $i\in \Omega$.

We denote with $p_i(t)$ the probability of the system to be in state $i$ at time t. These probabilities are nonnegative and normalized, $\bp \in [0,1]^N \,\text{ with } \| \bp \|_1 = 1$.

The Master equation that describes the change in time of these probabilities is the following initial value problem 
\begin{equation} \label{MasterEquation_components}
\begin{aligned}
\partial_t p_i(t) &= \mathlarger{\sum}\limits_{j=1\atop j\neq i}^N \; \big(\,  p_j(t) \, \gamma_{j\to i} - p_i(t) \, \gamma_{i \to j} \, \big) \\
p_i(t=0) &= p^{(i)}_0, 
\end{aligned}
\end{equation}
or, in matrix-vector notation,
\begin{equation}   \label{MasterEquation}
\begin{aligned}
\dot{\bp} &= \,\G \, \bp  \\
\bp(t=0) &= \bp_0 , \text{  with }
 \G_{ij}= \, \begin{cases}
 \g_{j\to i}  & ,  i \neq j \\
  -\mathlarger{\sum}\limits_{k=1}^N \g_{j\to k} &  , i = j\, .
\end{cases}
\end{aligned}
\end{equation}

The solution is given by $\bp_t := \e^{\G\, t} \, \bp_0$, with the initial state $\bp(t=0) = \bp_0$ and the solution operator
\begin{equation} \label{SolutionOperator}
\begin{aligned}
\e^{\G\, t} := \mathlarger{\sum}\limits_{k\in \N_0} \frac{\G^k \, t^k}{k!} = \lim\limits_{n\to \infty} \left( \Id + \frac{\G \, t}{n} \right)^n. 
\end{aligned}
\end{equation}

The matrix $\G$ is called the \textbf{generator} of the network $\Omega$ \cite{bremaud2013markov}. 

In the following, we relate properties of the solution $\bp_t := \e^{\G\, t} \, \bp_0$ to the properties of the network $\Omega$.  

Since the column sum of $\G$ is zero  $\left(\sum\limits_{i=1}^N \G_{ij} = 0\right)$, the matrix $\G$ has at least one eigenvalue 0. The eigenvectors to this eigenvalue are stationary states that satisfy $\dot \bp = 0$.  
From Gershgorin's circle theorem  \cite{huppert2006lineare} follows that all non-zero eigenvalues of $\G$ have a (strictly) negative real part.  

The column sum of any matrix power of $\G$ is also zero, $\sum\limits_{i=1}^N \left(\G^k\right)_{ij} = 0$ for all $k\in \N$, as can be shown by induction. 
Further, a power series expansion yields that the column sums of its matrix exponential $\e^{\G \, t}$ equals one, 

\begin{equation}\label{summehochgammat}
    \sum\limits_{i=1}^N \left(\e^{\G \, t}\right)_{ij} = 1.
\end{equation}

Every initial state can be written as a linear combination of generalized eigenvectors of $\G$. In the limit $t\to\infty$, only the eigenvectors of the eigenvalue $\lambda=0$ ( the ones lying in the kernel of $\G$) remain, while the contributions of the other eigenvectors decay exponentially.

To see this, let $\bv_{\lambda, m}$ be a generalized eigenvector of rank $m\in \N$ to the eigenvalue $\lambda$. The action of the time evolution operator $\e^{\G \, t}$ on this vector is
\begin{equation}\label{egammatv}
\begin{aligned}
\e^{\Gamma\, t}\, \bv_{\lambda, m} = \e^{\lambda\, t}\, \sum\limits_{k=0}^{m-1} \frac{t^k}{k!} \, \bv_{\lambda, m-k}\, .
\end{aligned}
\end{equation}

 For non-zero eigenvalues $\lambda \neq 0$, this tends to zero since  $ \text{Re}[\lambda] \, < \, 0 $. For $\lambda=0$, we must have $m=1$ since the solution of the master equation \eqref{MasterEquation} is bounded. This means that all generalized eigenvectors to the eigenvalue $\lambda=0$ lie in the kernel of $\G$.

 This results in 
 \begin{equation}
\begin{aligned}
\e^{\Gamma\, t}\, \bv_{\lambda=0} =  \bv_{\lambda=0}\, .
\end{aligned}
\end{equation}

\section{Definitions and preparatory considerations}
Before we can state and prove the main theorem, we need a few more definitions and theorems.

\subsection{Paths}
If state $b$ can be reached from state $a$ via a series of transitions, there is a {path} from $a$ to $b$, which we indicate by $a \rightsquigarrow b$, and we say that state $b$ is reachable from state $a$.  
We denote with
\begin{itemize}
    \item $\mathcal{R}(^{\rightarrow}a) := \{ b \in \Omega \,|\, b \rightsquigarrow a \}$ the set of states from where a path \textbf{to} $a$ exists and with
    \item $\mathcal{R}(a^{\rightarrow}) := \{ b \in \Omega \,|\, a \rightsquigarrow b \}$ the set of states to which a path \textbf{from} $a$ exists. 
\end{itemize}

\subsection{Weak, unilateral, and strong connectedness}

We call $\Omega$  
\begin{itemize}
\item[i)]  \textbf{weakly} connected if the corresponding undirected graph of $\Omega$  is connected; 
\item[ii)] \textbf{unilaterally} connected if for all $a, \, b \in \Omega$, $b$ is reachable from $a$ OR $a$ is reachable from $b$; 
\item[iii)] \textbf{strongly} connected if for all $a, \, b \in \Omega$, $b$ is reachable from $a$ and $a$ is reachable from $b$, \\
    $a \rightsquigarrow b \text{ {AND} } b \rightsquigarrow a$.  
    \end{itemize}
    Figure 1 illustrates the difference.
    \begin{figure}[H]
\begin{center}
\begin{subfigure}{0.35\textwidth}
  \subcaption{weakly, but not unilaterally connected}
  \label{Fig_1_b}
  \includegraphics[width=1.0\columnwidth]{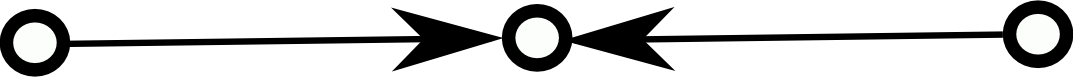} 
  \end{subfigure}\hspace*{15mm}\begin{subfigure}{0.35\textwidth}
  \subcaption{unilaterally, but not strongly connected}
  \label{Fig_1_c}
  \includegraphics[width=1.0\columnwidth]{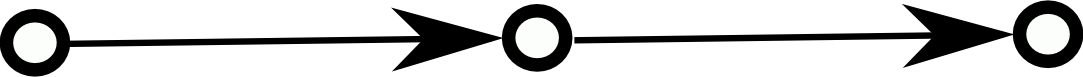} 
\end{subfigure}
\caption{The differences between weak, unilateral and strong connectivity. }
\end{center}
\end{figure}
\subsection{Relaxing networks}
We call $\Omega$ \textbf{relaxing} if there exists a unique stationary state $\bp_\infty $ such that for all initial conditions $\bp_0$ the dynamics of the Master equation converges to this  stationary state, that is $\lim\limits_{t\to\infty} \bp_t = \bp_\infty$. This means that the eigenspace of  $\G$ corresponding to the eigenvalue $0$ is one-dimensional,  $\text{dim}\left(\text{\,kern } (\G)\right) = 1$. 

\subsection{Absorbing subsets} \label{PropertiesOfReaching} 

We write $\bp \in B$ if all mass of the probability distribution is restricted to a subset $B\in\Omega$, that is $\mathlarger{\sum}\limits_{i \in B} p_i = 1$. 
We call a subset $B \subseteq \Omega$ {absorbing} if there are no edges pointing out of $B$, that is if $\g_{i\to j} = 0$ for all $i\in B$ and $j \in B^C$.
This means that probability cannot flow out of $B$, so when the mass of the probability distribution of the initial state is in $B$ ($\bp_0 \in B$), then it will stay in $B$ for all times: $\bp_t \in B$ for all $t\geq0$. 

Later, we will use the fact that $\mathcal{R}(^{\rightarrow}a)^C$ and $\mathcal{R}(a^{\rightarrow})$ are absorbing subsets. 
The two statements are shown as follows:
\begin{itemize}
\item If $\mathcal{R}(^{\rightarrow}a)^C$ were not absorbing, there would be a state $c \in \mathcal{R}(^{\rightarrow}a)^C$ and a $ b\in \mathcal{R}(^{\rightarrow}a)$ such that  $c \rightsquigarrow b$. However, this would imply $ c \rightsquigarrow b \rightsquigarrow a  $, which is a contradiction. 

\item If $\mathcal{R}(a^{\rightarrow})$ were not absorbing, there would be a state $b \in \mathcal{R}(a^{\rightarrow})$ and a state $c \in \mathcal{R}(a^{\rightarrow})^C$ such that  $b \rightsquigarrow c$. However, this would imply $ a \rightsquigarrow b \rightsquigarrow c  $, which is a contradiction. 
\end{itemize}

\subsection{Minimal absorbing subsets} \label{MinimalAbsorbingSubSets}
An absorbing subset $B\subseteq \Omega $ is called \textbf{minimal} if 
for all absorbing subsets $C\subseteq \Omega$ with $C \subseteq B$, we have $B = C. $
In particular, there can be more than one minimal absorbing subset.

Every minimal absorbing subset $B$ is strongly connected. To see this, assume that there are two states $i,\, j \in B$ with $i \nrightsquigarrow j$. 
Then $ i \in \left( \mathcal{R}(^\rightarrow j)^C \cap \, B \right)$ and $ j \notin \left( \mathcal{R}(^\rightarrow j)^C \cap \, B \right)$. This implies that  $\left( \mathcal{R}(^\rightarrow j)^C \cap \, B \right)$ is a non-empty intersection of two minimal absorbing sets which is strictly less than $B$, in contradiction with the premise that $B$ is minimal.

Every state $\omega \in \Omega$ leads to a (not necessarily unique) minimal absorbing subnetwork $B_\omega$,  that is there exists a path $\omega\to b$ to some element $b\in B_\omega$ of some minimal absorbing subset $B_\omega$. This follows from the result in the previous subsection that the set $\mathcal{R}(\omega^{\rightarrow})$ is absorbing. This set must contain a minimal absorbing set $B_\omega\subseteq \mathcal{R}(\omega^{\rightarrow})$, and  (since minimal absorbing subnetworks are strongly connected) every state of $B_\omega$ is reachable from $\omega$, that is $\omega \rightsquigarrow b$ for all states $b \in B_\omega \subseteq \mathcal{R}(\omega^{\rightarrow})$. 

\subsection{Diagonal dominance} \label{DiagonalDominance}
Let  $B \in \C^{N\times N}$ be a complex matrix. 
\begin{itemize}
    \item For $i \in \{ 1, \dots, N \}$, we call the $i$-th row of $B$ \textbf{strictly diagonal dominant} (SDD) if 
    $ | B_{ii} | > \sum\limits_{j=1, \atop j \neq i}^N | B_{ij} | $. 
    \item We call the matrix $B$ \textbf{strictly diagonal dominant} (SDD) if every row of $B$ is SDD. Due to the Gershgorin circle theorem  \cite{huppert2006lineare}, a SDD matrix is non-singular.
    \item The definition of a \textbf{weakly diagonal dominant} (WDD) matrix is the same as the previous one, but with a "$\geq$" sign instead of a ">" sign. 
    \item We call $B$ \textbf{weakly chained diagonal dominant} (WCDD) if $B$ is WDD and for all rows $i \in \{1, \dots, N\}$ that are not (SSD), there exists a $k \in \{1, \dots, N\}$ and a path $i=i_0 \to \dots \to i_k = j$ in the directed graph associated to $B$ to a (SSD) row $j \in \{1, \dots, N\}$ of $B$. 
\end{itemize}
It can be shown that WCDD matrices are non-singular ( see section \eqref{WCDD_MatricesNonSingular} in the Appendix for a proof). 

\subsection{Structure of the matrix  $\G$}
Below, we will use the matrix $\Gamma$ in a specific form that can be obtained by re-numbering the states. 

Let $\mathcal{B} = \{B_1, \dots, B_n\}, \, n\leq N $ be the set of minimal absorbing subsets and define $M:= N - \sum\limits_{i=1}^n |B_i|$. 
We number the states as follows: The first $M$ states are those not contained in minimal absorbing networks. Then we count the states which lie in minimal absorbing networks block-wise, that is
\begin{equation}
\begin{aligned}
{B_0} &:=\Omega\backslash \bigcup\limits_{i=1}^n B_i = \, \{1, \dots, M\} \\
B_i &= \{ M + \mathlarger{\sum}\limits_{j=1}^{i-1}|B_j| + 1, \dots, M + \mathlarger{\sum}\limits_{j=1}^{i}|B_j| \, \}, {\text{ for } i\in\{1, \dotsc, n\}}. 
\end{aligned}
\end{equation}

After this re-numbering of the states, we can write $\G$ in the following form

\begin{equation} \label{Block_Form_of_G}
\begin{aligned}
\G &=
\begin{pmatrix}
\G_{B_0} & \zero^{M\times |B_1|}        &  \dots &  \zero^{M\times |B_n|} \\ 
\G_{B_0 \to B_1}  & \G_{B_1} &        & \zero^{|B_1|\times |B_1|}   \\
\vdots &          & \ddots &   \\
\G_{B_0 \to B_n}  & \zero^{|B_n|\times |B_1|}        &        & \G_{B_n}
\end{pmatrix} 
    \text{ with the matrices } \\ 
\G_{B_0} \in \R^{M\times M}, \, 
\G_{B_i} &\in \R^{|B_i| \times |B_i|}
 \text{ and }  
\G_{B_0 \to B_i} \in \R^{(N-M) \times M}, 
\text{ for } i\in \{1, \dotsc, n\}
\end{aligned}
\end{equation}

If there is no matrix $\G_{B_0}$ $ (M=0) $ and there is only one minimal absorbing set $(n=1)$, then $\G$ is called \textbf{irreducible} \cite{huppert2006lineare,horn2012matrix}, otherwise it is called \textbf{reducible}. 

\begin{figure}[H]
\begin{center}
  \includegraphics[width=0.5\columnwidth]{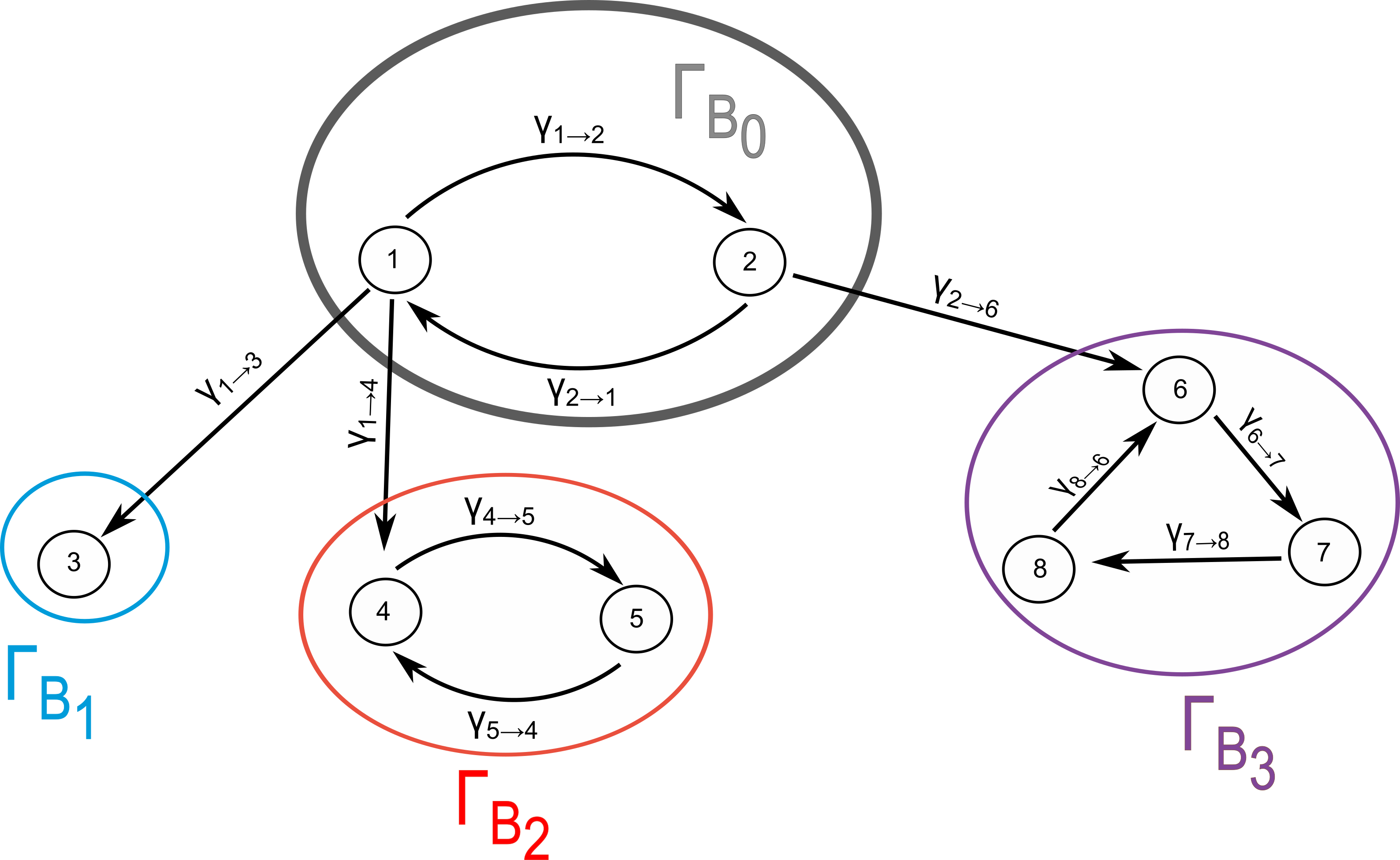} 
\caption{Example of a network / directed graph with the minimal absorbing sets being $B_1 = \{3\}$, $B_2 = \{4, \, 5\}$ and $B_3 = \{6, \, 7, \, 8\}$.   }
\label{Example_Network_Large} 
\end{center}
 \end{figure}

Figure~\ref{Example_Network_Large} gives an illustrating example. 

The matrices of the absorbing subnetworks and the full matrix $\Gamma$ for this example are
\begin{equation*}
\begin{aligned}
{\color{TUD_2b} \G_{B_1} = 0 }, \;   {\color{TUD_9b} \G_{B_2} = \begin{pmatrix} -\g_{4 \to 5} & \hspace*{3mm}\g_{5 \to 4} \\ \hspace*{3mm}\g_{4 \to 5} & -\g_{5 \to 4} \end{pmatrix} }, \;  {\color{TUD_11a} \G_{B_3} = \begin{pmatrix} -\g_{6 \to 7} &\hspace*{2mm} 0 & \hspace*{3mm}\g_{8 \to 6}  \\ \hspace*{3mm}\g_{6 \to 7} & -\g_{7 \to 8} & \hspace*{2mm}0 \\ \hspace*{2mm}0 & \hspace*{3mm}\g_{7 \to 8} & -\g_{8 \to 6} \end{pmatrix} }
\end{aligned}
\end{equation*}

and
\begin{align*}
\G =
\begin{pmatrix}
\G_{B_0}  & 0                           & 0                           &  0 \\ 
\G_{B_0\to B_1}   & {\color{TUD_2b} \G_{B_1} }  & 0                           & 0  \\
\G_{B_0\to B_2}   & 0                           & {\color{TUD_9b} \G_{B_2} }  & 0   \\
\G_{B_0\to B_3}   & 0                           &  0                          & {\color{TUD_11a} \G_{B_3} }
    \end{pmatrix} = 
 \begin{pmatrix}
-\g_{1\to 2}-\g_{1\to 3}-\g_{1\to 4} & \g_{2\to 1} & 0 & 0 & 0 & 0 & 0 & 0 \\
\g_{1\to 2} & -\g_{2\to 1}-\g_{2\to 6} & 0 & 0 & 0 & 0 & 0 & 0 \\
\g_{1\to 3} & 0  & {\color{TUD_2b}0} & 0 & 0 & 0 & 0 & 0 \\
\g_{1\to 4} & 0  & 0 & {\color{TUD_9b}-\g_{5\to 4}} & {\color{TUD_9b}\g_{4\to 5}} & 0 & 0 & 0 \\
0 & 0  & 0 & {\color{TUD_9b}\g_{5\to 4}} & {\color{TUD_9b}-\g_{4\to 5}} & 0 & 0 & 0 \\
0 & \g_{2\to 6}  & 0 & 0 & 0 & {\color{TUD_11a}-\g_{6\to 7}} & {\color{TUD_11a}0} & {\color{TUD_11a}\g_{8\to 6}} \\
0 & 0  & 0 & 0 & 0 & {\color{TUD_11a}\g_{6\to 7}} & {\color{TUD_11a}-\g_{7\to 8}} & {\color{TUD_11a}0} \\
0 & 0  & 0 & 0 & 0 & {\color{TUD_11a}0} & {\color{TUD_11a}\g_{7\to 8}} & {\color{TUD_11a}-\g_{8\to 6}} \\
\end{pmatrix}
\end{align*}


\subsection{If the matrix $\G_{B_0}$ exists, it is invertible } \label{Matrix_G_B_0_is_invertible}
In order to show that the matrix $\G_{B_0}$ is invertible, it suffices to show that $\G_{B_0}$ is WCDD (see section \eqref{WCDD_MatricesNonSingular} in the Appendix). 
We assume that $\G$ is of the form \eqref{Block_Form_of_G} and $\B = \{B_1, \dotsc, B_n\}$ is the set of minimal absorbing subsets. 
\begin{itemize}
 \item[1)] $\G_{B_0}$ is WDD, since  \\
\[
\sum\limits_{i=1, \atop i\neq j}^M  \underbrace{ |\left( \G_{B_0}\right)_{ij}| }_{ |\G_{ij}| } \leq \sum\limits_{i=1, \atop i\neq j}^N |\, \G_{ij} \, | = |\,\G_{jj}\,| = |\, \left( \G_{B_0}\right)_{jj}\,|. 
\]
\item[2)] Let $x \in B_0$  be an arbitrary state in $B_0$. Then there are two cases: 
\begin{itemize} 
 \item[ i) ] 
 If  $\g_{x \to b } > 0$  for some $b\in B_i$, then the  $x$-th row is SDD, since 
 \[ \sum\limits_{i=1, \atop i\neq x}^M \underbrace{ |\left( \G_{B_0}\right)_{i, \, x}| }_{ |\G_{i, \, x}| } < \sum\limits_{i=1, \atop i\neq x}^M  |\G_{i, \, x}| +  \underbrace{|\G_{b, \, x}|}_{\g_{x\to b}} \leq \sum\limits_{i=1, \atop i\neq x}^N  |\G_{i, \, x}| = |\G_{x, \, x}| = |\left( \G_{B_0}\right)_{x, \, x}|.   \]
 \item[ ii) ]
 If the $x$-th column is not SSD $(\g_{x\to b} = 0$ for all $b \in \Omega\backslash B_0$, we know from section \eqref{MinimalAbsorbingSubSets} that there exists a path from the state $x$ to a minimal absorbing set $B_i$, that is $x \to \dotsc \to x' \to b$ for some $i \in \{1, \dotsc, n\}$ and some state $b\in B_i$ and $ x' \in B_0$. But this again means that if the $x$-column is not SDD, there is a path from $x$ to the SDD column $x'$ (where $\g_{x'\to b}>0$, for some $b\in B_i$).  
 \end{itemize}
Hence, we conclude that $\G_{B_0}$ is WCDD and therefore invertible. 
\end{itemize}

\section{A simplified version of the main theorem}
Before stating the general condition for a network $\Omega$ to be relaxing, we first consider the important special case that $\Omega$ is strongly connected. Our goal in this section is to prove that a strongly connected network is relaxing (which is a corollary of the theorem of Perron-Frobenius  \cite{huppert2006lineare}). 
We proceed in three steps.

\subsection{A directed graph  $\Omega$ is strongly connected if and only if its adjacency matrix  $A$ is irreducible} \label{Strongly_Connected_iff_A_irreducible}

Here, we use a different but equivalent characterization of irreducibility than above:
A matrix $A \in \R^{N\times N}$ with non-negative entries $ (A_{ij} \geq \, 0 \text{ for all } i, j) $ is irreducible if and only if for all $i, j \in \{1, \dots, N\}$ there exists a natural number $k = k_{ij} \in \N $ 
such that the $i$-$j$-th entry of the  $k_{ij}$-th matrix power of $A$ is strictly greater zero, that is $ \left( A^{k_{ij}} \right)_{ij} > 0$. 

The adjacency matrix $A$ of the directed graph $\Omega$ is defined  by
\begin{equation}
\begin{aligned}
 A_{ij}= \, \begin{cases}
 1  & \g_{j\to i} > 0\\
 0   &  \text{else}. 
\end{cases} 
\end{aligned}
\end{equation}

In contrast to $\G$, the adjacency matrix $A$ has zeros on the main diagonal and only tells us \textit{qualitatively}, whether two links $j$ and $i$ are directly connected, and contains no \textit{quantitative} information about the strength of the links . 

 \begin{figure}[H]
\begin{center}
  \includegraphics[width=0.2\columnwidth]{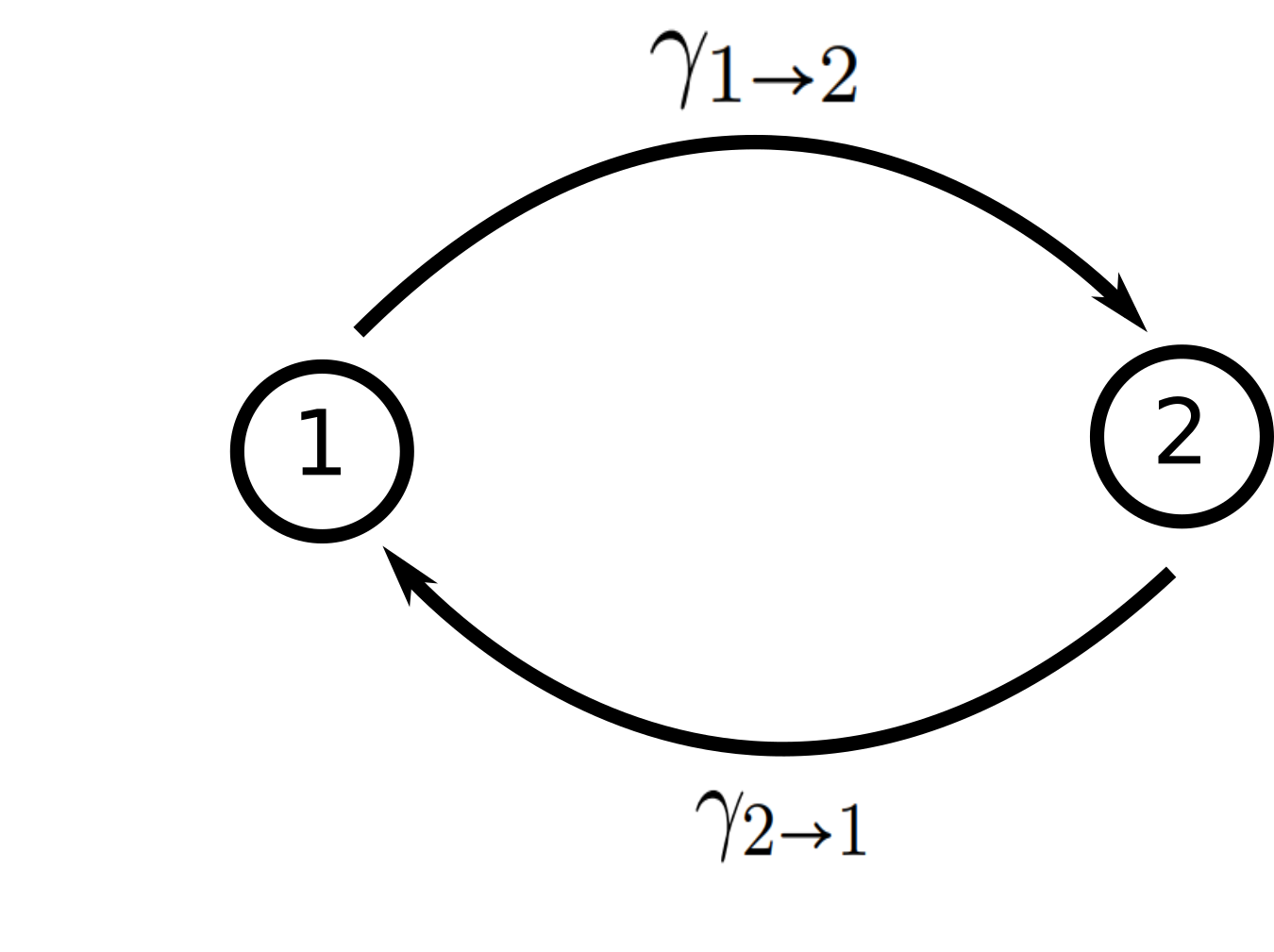} 
\caption{Illustrating the difference between the adjacency matrix $A$ and $\G$: 
}
\label{Difference_A_Gamma} 
\end{center}
 \end{figure}

Figure~\ref{Difference_A_Gamma} shows an example of a strongly connected network. The  corresponding  adjacency matrix is
$A = \begin{pmatrix}
0 & 1 \\
1 & 0
\end{pmatrix},$ and the generator is
$\G = \begin{pmatrix}
-\g_{1\to 2}  & \g_{2\to 1} \\
\g_{1\to 2} & -\g_{2\to 1}
\end{pmatrix}$. Unlike the adjacency matrix $A$, all the column sums of the generator $\G$ equal zero. 

The statement is proven by showing that a state $i \in \{1, \dots, N\} $ is reachable from another state $j \in \{1, \dots, N\}$ in $k_{ij}\in \N$ steps if and only if $\left(A^{k_{ij}}\right)_{ij} >0$, which can be proven by induction. 

Below, we will make use of a modified version of this result: We define $\tilde{A} := A + \Id$, which is the adjacency matrix of the directed graph $\tilde{\Omega}$ that is obtained by adding a self-loop to every state of $\Omega$.  Then $\tilde\Omega$ (and therefore also $\Omega$) is strongly connected if and only if $\tilde{A} $ is irreducible. 
But when $\tilde{\Omega}$ is strongly connected and every state has a self-loop, then there exists a natural number $k\in \N$ such that there is a path of length $n\in\N$ from state $j$ to state $i$ for all $n\geq k$. Hence, we can choose $k$ (defined at the beginning of section \eqref{Strongly_Connected_iff_A_irreducible}) to be independent of $i$ and $j$. In particular, when $\Omega$ is strongly connected the matrix power of the adjacency matrix $\tilde{A}^k = \left(\Id + A\right)^k$ has strictly positive entries when $k$ is large enough.  

This does not necessarily apply to the adjacency matrix $A$, as is obvious from the example  of Figure~\ref{Difference_A_Gamma}  with 
$A=\begin{pmatrix} 0 & 1 \\ 1 & 0\end{pmatrix}$. Here, the sequence $\left(A^m\right)_{m\in\N}$ of matrix powers of $A$ oscillates between the two matrices $ A^{2\,m} = \Id $ and $ A^{2\,m+1} = A $.

\subsection{When $\Omega$ is strongly connected, then for times $t>0$ all entries of the solution operator $\e^{\G \, t}$ are strictly positive, that is $ \left( \e^{\G \, t} \right)_{ij} > 0$, for all $i, \, j \in \{1, \dots, N\}$, where $\G$ is the corresponding generator of the network $\Omega$}. 

We know the following two statements to be true:  \\
Firstly, since $\Omega$ is strongly connected, the entries of $\left( \Id + \text{Adj}(\Omega)\right)^n$ are strictly positive when $n$ is large enough. \\
Secondly, the entry of $\left( \Id + \text{Adj}(\Omega)\right)_{ij}$ is strictly positive if and only if the entry of
\begin{equation*}
\begin{aligned}
\left( \Id + \frac{\G \, t}{n} \right)_{ij} \xlongequal{\eqref{MasterEquation}}
\begin{cases}
\, & 1 -  \frac{t}{n} \, \sum\limits_{m=1 \atop m\neq j}^N \g_{j\to m}
 \text{  , if } i=j  \\
\, &\frac{t}{n} \, \g_{j \to i} \hspace*{14mm}  \text{ , if } i\neq j, 
\end{cases}
\end{aligned}
\end{equation*}

is strictly positive for large enough $n$. \\

Hence, the entries of $\left( \Id + \frac{\G \, t}{n} \right)^n$ are strictly positive when $n$ is large enough. \\
Since by lemma \eqref{e_Gamme_i_j_LowerBound} (see Appendix)
the sequence $\left( \left[\left( \Id + \frac{\G \, t}{n} \right)^n\right]_{ij} \right)_{n\in \N}$ has a strictly positive lower bound, we conclude that the components of the solution operator $\left(\e^{\G \, t}\right)_{ij} := \lim\limits_{n\to\infty} \left[\left( \Id + \frac{\G \, t}{n} \right)^n\right]_{ij}$ are strictly positive. \\ 

The $ij$-th entry of $\left( \e^{\G \, t} \right) $ can be interpreted as the probability for being in state $i$ after the time $t>0$, provided that the system was in state $j$ at time $t=0$. 
So when all entries of the matrix $ \left( \e^{\G \, t} \right) $ are strictly positive, after an arbitrary small time $t>0$ every state has a non-zero probability (probability vectors $\bp_t$  are strictly positive), independent of the initial condition. \\

\subsection{When $\Omega$ is strongly connected, then $\Omega$ is relaxing and all components of the unique steady state are strictly positive,  $\bp_\infty \in \left(\R_{>0}\right)^N$ .} \label{Strongly_connected_implies_relaxing}

\begin{proof}
Let $\bv$ be an eigenvector of $\G$ to a real eigenvalue $\lambda \in \R$, that is $\G \, \bv = \lambda \, \bv$. Then we have
\begin{align*}
\e^{\lambda \, t} \|\bv\|_1 =&  \|\e^{\lambda \, t} \bv\|_1 = \|\e^{\G \, t} \bv\|_1 = \sum\limits_{i=1}^N \left|\;  \sum\limits_{j=1}^N \left( \e^{\G \, t} \right)_{ij} \, v_j \; \right| \overset{{\color{ElectricPurple} (*)}} \leq \sum\limits_{i,\, j=1}^N
 \left| \, \underbrace{\left( \e^{\G \, t} \right)_{ij}}_{\geq \, 0} \, \right| \cdot |v_j|  =
 \sum\limits_{j=1}^N |v_j| \; \underbrace{\left(\sum\limits_{i=1}^N \left( \e^{\G \, t} \right)_{ij}\right)}_{=\,1} = \| \bv \|_1. 
 \end{align*}
 The identity $\sum\limits_{i=1}^N \left(\e^{\G \, t} \right)_{ij} = 1$ was obtained above, equation \eqref{summehochgammat}.
 Now, let $\bv \in \text{kern } (\G) $, which implies $\lambda = 0$ and we have equality in the above estimation. But on the other hand, we have equality  in  {\color{ElectricPurple} (*)} if and only if
 
 \begin{equation}
\begin{aligned}\label{CrucialCondition}
\Bigl(\left( \e^{\G \, t} \right)_{i 1} \, v_1 , \dotsc,  \left( \e^{\G \, t} \right)_{i N} \, v_N\Bigr) \in \left(\R_{\geq \, 0} \right)^N \cup \left(\R_{\leq \, 0} \right)^N. 
\end{aligned}
 \end{equation}

 Since we assumed that $\Omega$ is strongly connected, we know from the previous subsection that the solution operator $\e^{\G \, t}$ has only strictly positive entries, so from equation \eqref{CrucialCondition} follows that $\colvec{3}{v_1}{\vdots}{v_N} = \bv \in \left(\R_{\geq \, 0} \right)^N \cup \left(\R_{\leq \, 0} \right)^N$. 
 
 This means that $\text{kern } (\G) \subseteq \left(\R_{\geq \, 0} \right)^N \cup \left(\R_{\leq \, 0} \right)^N$. But a vector space that contains only vectors where all entries have the same sign must be one-dimensional. If it was more than one-dimensional, one could create a vector with positive and negative entries by building a suitable linear combination of the two basis vectors. \\ 
 This proves that $\Omega$ is relaxing. It remains to be shown that the steady has strictly positive entries. So let $\bp_\infty \in \Kern(\G) $ be the unique steady state, with $\|\bp_\infty\|=1$. Then $\bp_\infty$ has only strictly positive components since we have equality in \eqref{p_infty_i_strictly_positive} if and only if all components of $\bp_\infty$ are zero: 
 
\begin{equation} \label{p_infty_i_strictly_positive}
\begin{aligned}
p_\infty^{(i)} &= \frac{\sum\limits_{j=1, \atop j\neq i}^N p_\infty^{(j)} \, \g_{j\to i} }{\sum\limits_{j=1, \atop j\neq i}^N  \g_{i\to j}} \, \geq 0 \, .\\
\end{aligned}
\end{equation}
\end{proof}

\section{Statement of the main theorem and its proof}\label{MainTheorem_1}
The theorem that we will prove in the following states: 

The Master equation \eqref{MasterEquation} of a network $\Omega$ is relaxing if and only if
there is exactly one minimal absorbing set.

We prove the forward direction by contraposition. Suppose there exists two distinct minimal absorbing subsets  $B_1, B_2 \subseteq  \Omega$. Then $\Omega$ cannot be relaxing since the stationary state  will lie in $B_1$ $(B_2)$ when the initial condition is in $B_1$ $(B_2)$, that is $\bp_0 \in B_{1,2} \Longrightarrow \bp_\infty \in  B_{1,2}$. \\

For the backward direction, we start from the assumption that there is exactly one minimal absorbing set $B_1$. We show that $\Omega$ is relaxing by proving $\text{dim kern} \, (\G) = 1$. we write $\G$ in the form \eqref{Block_Form_of_G},  

\begin{equation*}
\begin{aligned}
\G = \begin{pmatrix}
\G_{B_0} & 0 \\
\G_{B_0 \to B_1} & \G_{B_1}
\end{pmatrix}.
\end{aligned}
\end{equation*}

Let 
$\bv^* = \colvec{2}{\bu^*}{\bw^*} \in \Kern  \,(\G)$
with $\bu^* \in \R^M$ and $\bw^* \in \R^{(N-M)}$. Then we have
\begin{equation}
\begin{aligned}
\zero \overset{!}= \G \boldsymbol{v^*} = \begin{pmatrix}
\G_{B_0} & 0 \\
\G_{B_0\to B_1} & \G_{B_1}
\end{pmatrix} \colvec{2}{\bu^*}{\bv^*} =  \colvec{2}{\G_{B_0} \,  \bu^*}{\G_{B_0\to B_1} \,  \bu^* + \G_{B_1} \, \boldsymbol{w^*}}. 
\end{aligned}
\end{equation}
Since we know from section \eqref{Matrix_G_B_0_is_invertible} that $\G_{B_0}$ is invertible, we have  $\bu^* = \zero$ and $\bv^*= \colvec{2}{\zero}{\bw^*}$. But $\bw^*$ must lie in the kernel of $\G_{B_1}$, which is one-dimensional, since $B_1$ is strongly connected (see: section \ref{Strongly_connected_implies_relaxing}). This means that $\text{dim kern} \,  (\G) = \text{dim kern} \,(\G_{B_1}) = 1$ which completes the proof.

\section{Generalized main theorem: The number of minimal absorbing sets equals the dimension of the kernel of $\G$ } \label{GeneralizedMainTheorem}
\begin{proof}

Let $\B = \{B_1, \dotsc, B_n\}, \, n \leq N$ be the set of minimal absorbing sets and let $\G$ be of the form \eqref{Block_Form_of_G}, that is 
\begin{equation}
\begin{aligned}
\G = 
\begin{pmatrix}
\G_{B_0} & 0        &  \dots &  0 \\ 
\G_{B_0 \to B_1}  & \G_{B_1} &        & 0  \\
\vdots &          & \ddots &   \\
\G_{B_0 \to B_n}  & 0        &        & \G_{B_n}
    \end{pmatrix}. 
\end{aligned}
\end{equation}
Let $\bq_i \in \Kern(\G_{B_i}) \cap \left( \R_{>0} \right)^{|B_i|}$ with $\| \bq_i\|_1=1$  for all  $i \in \{1, \dots, n\}$. 
We know from section \eqref{MinimalAbsorbingSubSets} that all $B_i$ are strongly connected and from section \eqref{Strongly_connected_implies_relaxing} that all $\bq_i$ are well defined and uniquely determined. 

Further, define $\bp_i := (\zero_M, \zero_{|B_1|}, \dotsc, \bq_i, \dotsc, \zero_{|B_n|})^T $ and note that the vectors $ \{ \bp_i \,:\, i \in \{1, \dotsc, n\} \} $ are linearly independent. 
Then we have 
\begin{equation}
\begin{aligned}
\G \, \bp_i = \colvec{3}{\zero}{\G_{B_i} \, \bq_i}{\zero} = \zero_N. 
\end{aligned}
\end{equation}
and hence $\Span \left( \{\bp_i \,:\, i \in \{1, \dotsc, n\}\} \right) \subseteq \Kern(\G) $. \\

On the other hand, let $\bv \in \Kern(\G)$ be an arbitrary element of the kernel of $\G$. We write $\bv$ as 
\begin{equation}
\begin{aligned}
\bv = (\bu_0, \bw_1, \dotsc, \bw_n) \in \R^M \times \R^{|B_1|} \times \dots \times \R^{|B_n|}.
\end{aligned}
\end{equation}

Then we know that

\begin{equation}
\begin{aligned}
\zero_N 
\xlongequal{v\, \in \,  \Kern(\G)} \G \, \bv =
\colvec{4}
{\G_{B_0} \, \bu_0}
{ \G_{B_0 \to B_1} \, \bu_0 + \G_{B_1} \, \bw_1 }
{\vdots}
{ \G_{B_0 \to B_n} \, \bu_0 + \G_{B_n} \, \bw_n }
= \colvec{4}
{\zero_M}
{  \G_{B_1} \, \bw_1 }
{\vdots}
{  \G_{B_n} \, \bw_n }. 
\end{aligned}
\end{equation}

The last equality holds, because $\G_{B_0}$ is invertible. Moreover, since $\text{dim } \Kern(\G_{B_i}) = 1$ for all $i \in \{1, \dotsc, n\}$ we conclude that $\bu_0 = \zero_M$ and $\bw_i \in \Kern(\G_{B_i}) = \Span(\bq_i)$, that is $\bw_i = \lambda_i \, \bq_i$, for some $\lambda \in \R$. 

It then follows that 

\begin{equation}
\begin{aligned}
\bv &= \colvec{4}{\zero_M}{\lambda_1 \, \bq_1}{\vdots}{\lambda_n \, \bq_n} = \sum\limits_{i=1}^n \lambda_i \, \bp_i, \text{ and hence} \\
\Kern(\G) &\subseteq \Span \Bigl( \bigl\{\bp_i \,:\, i \in \{1, \dotsc, n\} \bigr\} \Bigr). 
\end{aligned}
\end{equation}
\end{proof}

This means that we can construct a basis of steady states from the set of minimal absorbing sets, with every basis vector corresponding to exactly one minimal absorbing set.

For the example given in Figure~\ref{Example_Network_Large}, the basis vectors are 
\begin{equation}
\begin{aligned}
{\color{TUD_2b} \bp_1 = \colvec{8}{0}{0}{1}{0}{0}{0}{0}{0}
}, \, 
{\color{TUD_9b}\bp_2 = \frac{1}{\g_{4\to 5}+\g_{5\to 4}} \colvec{8}{0}{0}{0}{\g_{5\to 4}}{\g_{4\to 5}}{0}{0}{0}
}, \, 
{\color{TUD_11a} \bp_3 = \frac{1}{ \g_{7 \to 8}\, \g_{8 \to 6} + \g_{8 \to 6} \, \g_{6 \to 7} +\g_{6 \to 7} \, \g_{7 \to 8}}
\colvec{8}{0}{0}{0}{0}{0}{ \g_{7 \to 8}\, \g_{8 \to 6} }{\g_{8 \to 6} \, \g_{6 \to 7} }{\g_{6 \to 7} \, \g_{7 \to 8}}
}.
\end{aligned}
\end{equation}


\section{Discussion and Conclusions}

We have shown that the information about the number of linearly independent steady states of a finite-size Master equation is encoded in the directed network of transitions: There is a one-to-one correspondence between minimal absorbing sets and basis vectors of the kernel of the generator that span the space of steady states. In particular, the dimension of the space of steady states equals the number of minimal absorbing sets. Moreover, for every minimal absorbing set $B_i \subseteq \Omega$ we can construct a normalized basis vector $\bp_i \in \left(\R_{\geq\, 0}\right)^N \cap \Kern(\G)$, where the strictly positive entries correspond to the states lying in the minimal absorbing set $B_i$. 

The questions remains how to proceed in practice. Given a system $\s = (\Omega, \E)$ composed of a finite number of states $\Omega$ and transition rates $\E$, how does one determine whether this system is relaxing and, if not, how many steady states there are?

\begin{figure}[H]
\begin{center}
\begin{subfigure}{0.49\textwidth}
   \subcaption{original network}
      \label{MacroState_Network_OriginalNetwork}
  \includegraphics[width=0.99\columnwidth]{Example_Network_Large.png} 
\end{subfigure}  \begin{subfigure}{0.49\textwidth}
   \subcaption{macrostate network}
   \label{MacroState_Network_MacroStateNetwork}
  \hspace*{12mm}\includegraphics[width=0.6\columnwidth]{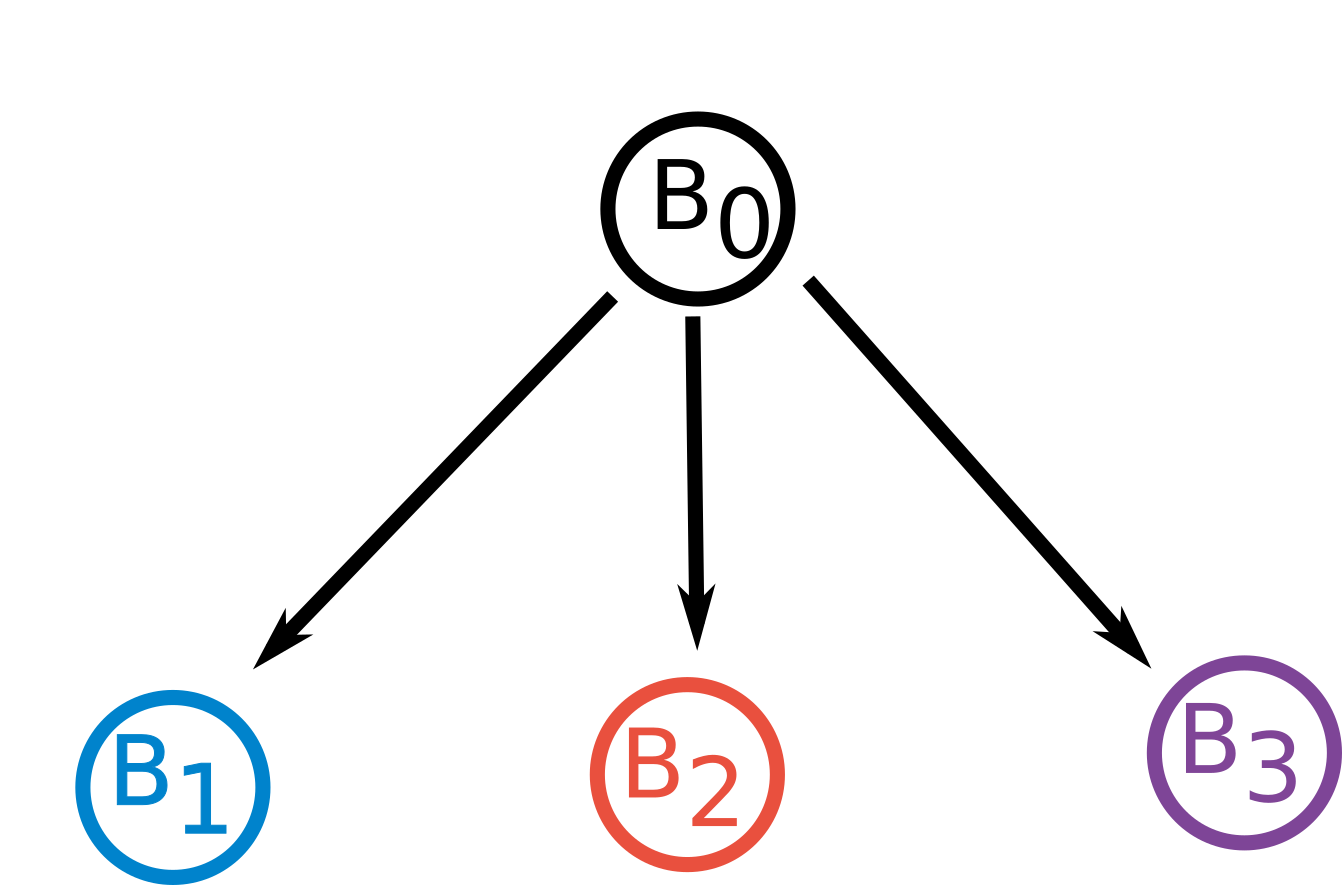} \end{subfigure}
\caption{Merging of the strongly connected components of a network into 'macrostates'.  }
\label{MacroState_Network} 
\end{center}
\end{figure}

The standard way would be to determine the dimension and the span of the kernel of $\G$, which can be done in $\Order(|\Omega|^3)$ steps. 
The alternative, however, is to search for the strongly connected components, which is possible in $\Order(|\Omega| + |\E|)$ steps, and merge each of these strongly connected components into a single `macrostate', see Figure~\ref{MacroState_Network} where this is done for the example from Figure~\ref{Example_Network_Large}. 

Every absorbing macrostate in the resulting coarse-grained network is a minimal absorbing set in the original network.  Their number determines whether $\Omega$ is relaxing or not. 

In order to find a basis of vectors spanning the space of steady states, it suffices now to determine the steady state $\bp_i$ for each minimal absorbing set $B_i$, based on equation \eqref{AnalyticalExpressionForStationaryStatesInMinimalAbsorbingSets}
in the Appendix. 

The stationary state $\bp_\infty$ for the whole network $\Omega$ when starting from the initial state $\bp_0$ is the projection of $\bp_0$ onto the span of the steady states of the minimal absorbing sets $\bp_i$. Writing the initial state as  

\begin{equation}
\begin{aligned}
\bp_0 = \sum\limits_{i=1}^n \lambda_i \, \bp_i + \text{rest}, 
\end{aligned}
\end{equation}
with the rest being a vector orthogonal to the $\bp_i$, 
then $\bp_\infty$ is given by $\bp_\infty = \sum\limits_{i=1}^n \lambda_i \, \bp_i $.

The proof provided in this paper is based on the assumption that the state space is finite.  The theorems on which the proof is based, such as the Perron-Frobenius theorem \cite{huppert2006lineare}), have no version for the infinite-dimensional case. 
There are, however, good reasons to assume that the results are valid also for some models with an infinite state space. 

There are systems which show the same behavior (qualitative and approximately quantitative) as a reduced system $\Omega_{< \,\infty} \subsetneqq \Omega_\infty $ with a finite state space, for which the above considerations hold. Examples are  systems with a finite number of minimal absorbing sets for each connected component and chemical reaction systems where an arbitrary large number of molecules is extremely unlikely. For these systems, we expect that given an initial state $\bp_0$ and an $\epsilon>0$,  it is possible to choose a finite sub-system $\Omega_{<\infty}$ of $\Omega_\infty$ such that the apart from an arbitrary small probability mass $\epsilon$ dynamics takes places in the finite sub-system.

\appendix
\section{Appendix}\label{Appendix}
\subsection{WCDD matrices are non-singular} \label{WCDD_MatricesNonSingular}  

\begin{proof}
The fact that SDD matrices are non-singular follows from Gershgorin's circle theorem \cite{huppert2006lineare}. Now let $A$ be WCDD and assume that $A$ is singular. Let $\boldsymbol{x} \in \text{kern} (A)$, w.l.o.g. assume that there is an $i \in \{1, \dots, N\}$ such that $1 = | x_i | \geq | x_k| \, \forall k \in \{1, \dots, N\}, \, k\neq i$. Then we have
\begin{align}
0 =& \,  (A \, \boldsymbol{x})_i = \sum\limits_{k=1, \atop k\neq i}^N A_{ik} \, x_k + A_{ii} \, x_i \Longrightarrow - A_{ii} \, x_i = \sum\limits_{k=1, \atop k\neq i}^N A_{ik} \, x_k \label{WCDD_1} \\
\Longrightarrow | A_{ii} | =& \, \left|\,  - A_{ii} \, x_i\,  \right| \overset{\eqref{WCDD_1}}= \left| \sum\limits_{k=1, \atop k\neq i}^N A_{ik} \, x_k \right | \leq  \sum\limits_{k=1, \atop k\neq i}^N |\,  A_{ik} \, | \underbrace{| \, x_k |}_{\leq 1} \overset{{\color{MyGreen}(*)}} \leq \sum\limits_{k=1, \atop k\neq i}^N |  A_{ik}| \overset{A\dots \, WDD} \leq | \,A_{ii} \, | \label{WCDD_2}
\end{align}
Hence, in line \eqref{WCDD_2} we have equality everywhere. In particular: 
\begin{itemize}
    \item[i)] The last equality tells us that the $i$-the row is not SDD. 
    \item[ii)] Equality  in {\color{MyGreen} (*)} tells us, that whenever $A_{ik} \neq \, 0 \Longrightarrow |x_k| = 1 $. 
\end{itemize}
Since $A$ is WCDD we know there exists a path $i=i_0 \to i_1 \to \dots \to i_k = j$ to the SDD row number $j$. In particular, we have $A_{i_0, \, i_1} \neq \, 0 \overset{ii)} \Longrightarrow |x_{i_1}| = 1 $. Repeating the argument from the beginning, we get from i) that the $i_1$-th row is not SDD. When we keep iterating, we finally get, that the $j$-th row is not SDD which is a contradiction. 
\end{proof}


\subsection{When $\Omega$ is strongly connected, the sequence $\left( \left[\left( \Id + \frac{\G \, t}{n} \right)^n\right]_{ij} \right)_{n\in \N}$ has a strictly positive lower bound} \label{e_Gamme_i_j_LowerBound}

\begin{proof}
Define 
\begin{equation}
\begin{aligned}
\G_\text{min} &:= \min \{ \G_{ll} \,:\, l \in \{1, \dots, N\}\} < 0 \\
d &:= \text{ length of the shortest path from $j$ to $i$  } \\
\g^{(j\to i)} &:= 
\begin{cases}
1, \text{ if $j=i$ } \\
\min \left\{ \prod\limits_{l=1}^d \g_{m_l \to m_{l+1}}  \,:\, (m_1, \dotsc, m_d) \in \{1, \dotsc,N \}^d, \, m_1=j, \, m_d=i, \, \g_{m_l \to m_{l+1}} \neq 0 \right\}, \text{ if $j\neq \, i$ }
\end{cases}
\end{aligned}
\end{equation}
The interpretation of $\g^{(j\to i)}$ is the following: 
For all path $(j=m_i \to  \dotsc \to m_k=i)$ of length $d\in\N$ from state $j$ to state $i$, multiply the rates $\g_{m_l \to m_{l+1}}$ of the edges that constitute the path and take the minimum over all paths. \\

Then we can make the following estimate:  
\begin{equation}
\begin{aligned}\label{Gamme_ij_FindingLowerBound}
\left( \e^{\G \, t} \right)_{ij} 
&= \lim\limits_{n\to \infty} 
\left(\left(\Id + \frac{\G \, t}{n} \right)^n\right)_{ij} = 
\lim\limits_{n\to \infty}
\sum\limits_{k_1=1}^N \cdots \sum\limits_{k_n=1}^N \left(\Id + \frac{\G \, t}{n} \right)_{i. \, k_1} \dotsc \; \left(\Id + \frac{\G \, t}{n} \right)_{k_{n-1}\, j}
\geq \\
& \overset{{\color{ElectricPurple}(*)}}\geq
\lim\limits_{n\to \infty}
\underbrace{\frac{n!}{(n-d)! \, n^d}}_{ \xlongrightarrow{n\to \infty} 1} \, 
 \, \frac{\g^{(j\to i)} \, t^d}{d!} \, 
 \underbrace{\left( 1+\frac{\G_{min} \, t}{n} \right)^{n-d}}_{
\xlongrightarrow{n\to \infty} \, \e^{\G_{min} \, t} }
= \,  \frac{\g^{(j\to i)} \, t^d}{d!} \, \e^{\G_{min} \, t} 
\,> \, 0, 
\end{aligned}
\end{equation}

where we used in {\color{ElectricPurple}(*)} the following calculation: 

\begin{equation}
\begin{aligned} \label{Gamme_ij_FindingLowerBound_AuxiliaryCalculation}
\sum\limits_{k_1=1}^N \cdots& \sum\limits_{k_{n-1}=1}^N \left(\Id + \frac{\G \, t}{n} \right)_{i \, k_1} \dotsc \; \left(\Id + \frac{\G \, t}{n} \right)_{k_{n-1}\, j} 
\geq \colvec{2}{n}{d} 
\underbrace{
\left(\frac{\g_{j\to m_2} \, t}{n} \right) \dotsc
\left(\frac{\g_{m_{d-1}\to i} \, t}{n} \right) 
}_{\text{ d times }} \; 
\underbrace{
\left(1 + \frac{\G_{\Box\Box} \, t}{n} \right) \dotsc
\left(1 + \frac{\G_{\Box\Box} \, t}{n} \right) 
}_{\text{ n-d times }} \geq \\ 
&\geq 
 \frac{n!}{(n-d)! \, n^d} \, 
 \, \frac{\g^{(j\to i)} \, t^d}{d!} \, 
 \left( 1+\frac{\G_{min} \, t}{n} \right)^{n-d}
\end{aligned}
\end{equation}
\end{proof}

\subsection{ Explicit expression for the  stationary state of  a strongly connected network }

\begin{figure}[H]
\begin{center}
  \includegraphics[width=0.3\columnwidth]{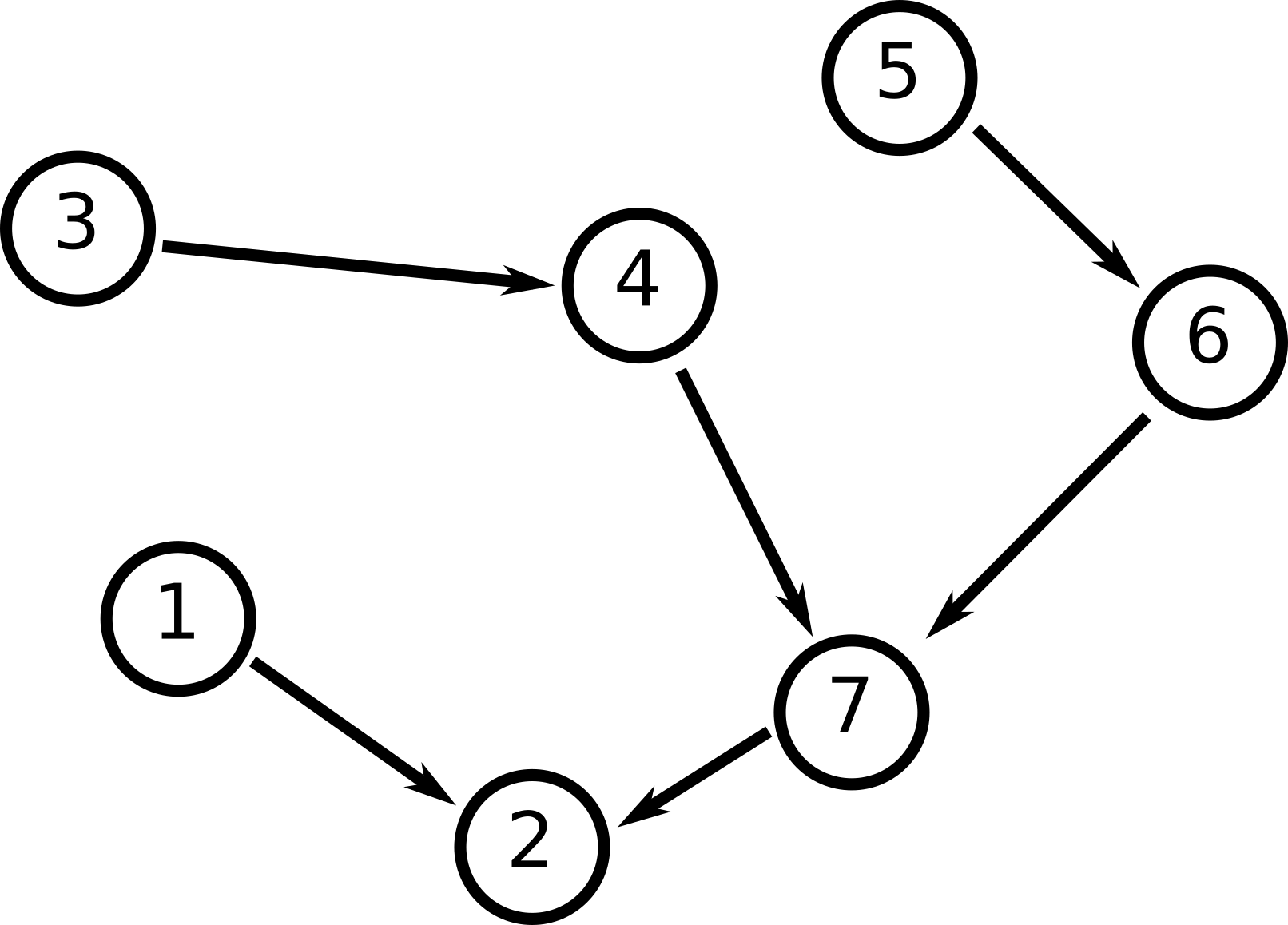} 
\caption{ Example of an in-tree rooted at state number $2$. There is a unique directed path from each state of the tree leading to state number $2$.  }
\label{Example_InTreeRootedAtState_2} 
\end{center}
\end{figure}

We call a network an \textbf{in-tree} (also called \textbf{anti-arborescence} \cite{gabow1978finding}) \textbf{rooted at state $\omega_0 \in \Omega$}, if for all states $\omega \in \Omega$ there is a unique directed path leading from state $\omega$ towards the root $\omega_0$. 
An example is given in Figure \ref{Example_InTreeRootedAtState_2}, where the root is state number $2$, the distinguished state to which all the directed paths lead. 

Further, we define 
\begin{equation}
\begin{aligned}
\Theta_m(\Omega) := \left\{ T \subseteq \s \,:\,  
T \text{ is an in-tree of $\Omega$ rooted in state number $m$ }  
\right\}
\end{aligned}
\end{equation}

to be the set of all in-tree of $\Omega$ that are rooted in state number $m \in \{1, \dotsc, N\}$. \\

Now, suppose that the network $\Omega$ is strongly connected. Then the kernel of the corresponding generator $\G$ is one-dimensional (see: section \ref{Strongly_connected_implies_relaxing}) and there is an analytical expression for it, namely: 

\begin{equation} \label{AnalyticalExpressionForStationaryStatesInMinimalAbsorbingSets}
\begin{aligned}
\Kern(\G) &= \Span \left\{  \left(\mathlarger{\mathlarger{\sum\limits}}_{T\, \in \, \Theta_m(\Omega)} \; \mathlarger{\mathlarger{\prod\limits}}_{(i,j) \, \in \,  \E(T)} \g_{i \to j} \right)_{m \in \{1, \dotsc, N\}} \right\} \\
&= \Span \left\{  \colvec{3}{
\mathlarger{\mathlarger{\sum\limits}}_{T\, \in \, \Theta_1(\Omega)} \; \mathlarger{\mathlarger{\prod\limits}}_{(i,j) \, \in \,  \E(T)} \g_{i \to j} 
}{\\ \vdots \\ }{
\mathlarger{\mathlarger{\sum\limits}}_{T\, \in \, \Theta_N(\Omega)} \; \mathlarger{\mathlarger{\prod\limits}}_{(i,j) \, \in \,  \E(T)} \g_{i \to j} 
}\right\}.  
\end{aligned}
\end{equation}

\begin{figure} 
\begin{minipage}{0.3\textwidth}
\begin{center}
\begin{subfigure}{1.0\textwidth}
\subcaption{}
\label{Example_StationaryState_OriginalNetwork}
  \includegraphics[width=0.8\columnwidth]{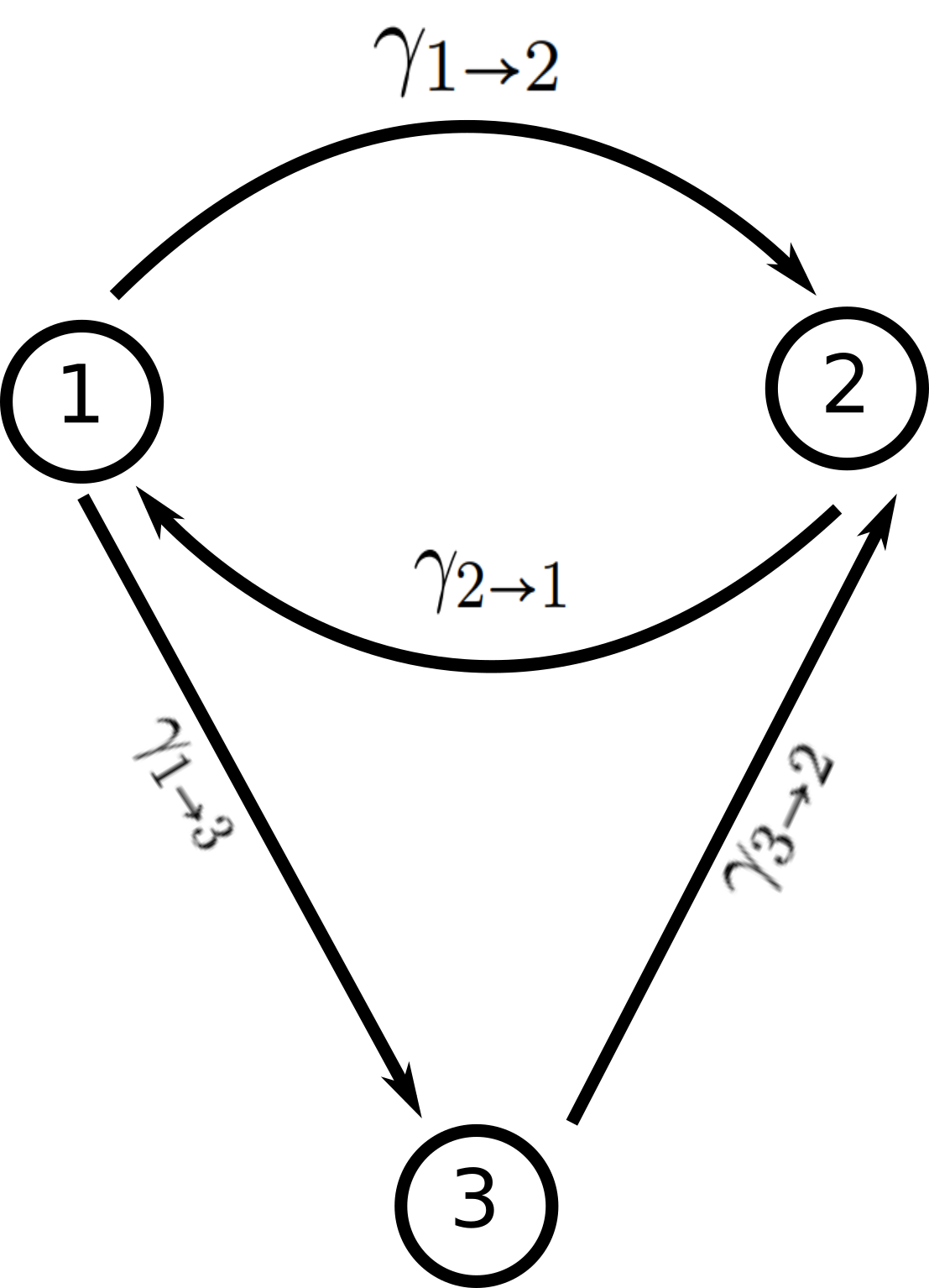}   
\end{subfigure}
\end{center}
\end{minipage} 
\hspace*{3mm} 
\begin{minipage}{0.7\textwidth}
\begin{center}
 \begin{subfigure}{0.25\textwidth}
\subcaption{In-tree rooted in $1$}
\label{Example_StationaryState_InTrees_LeadingTo_1}
  \includegraphics[width=1.0\columnwidth]{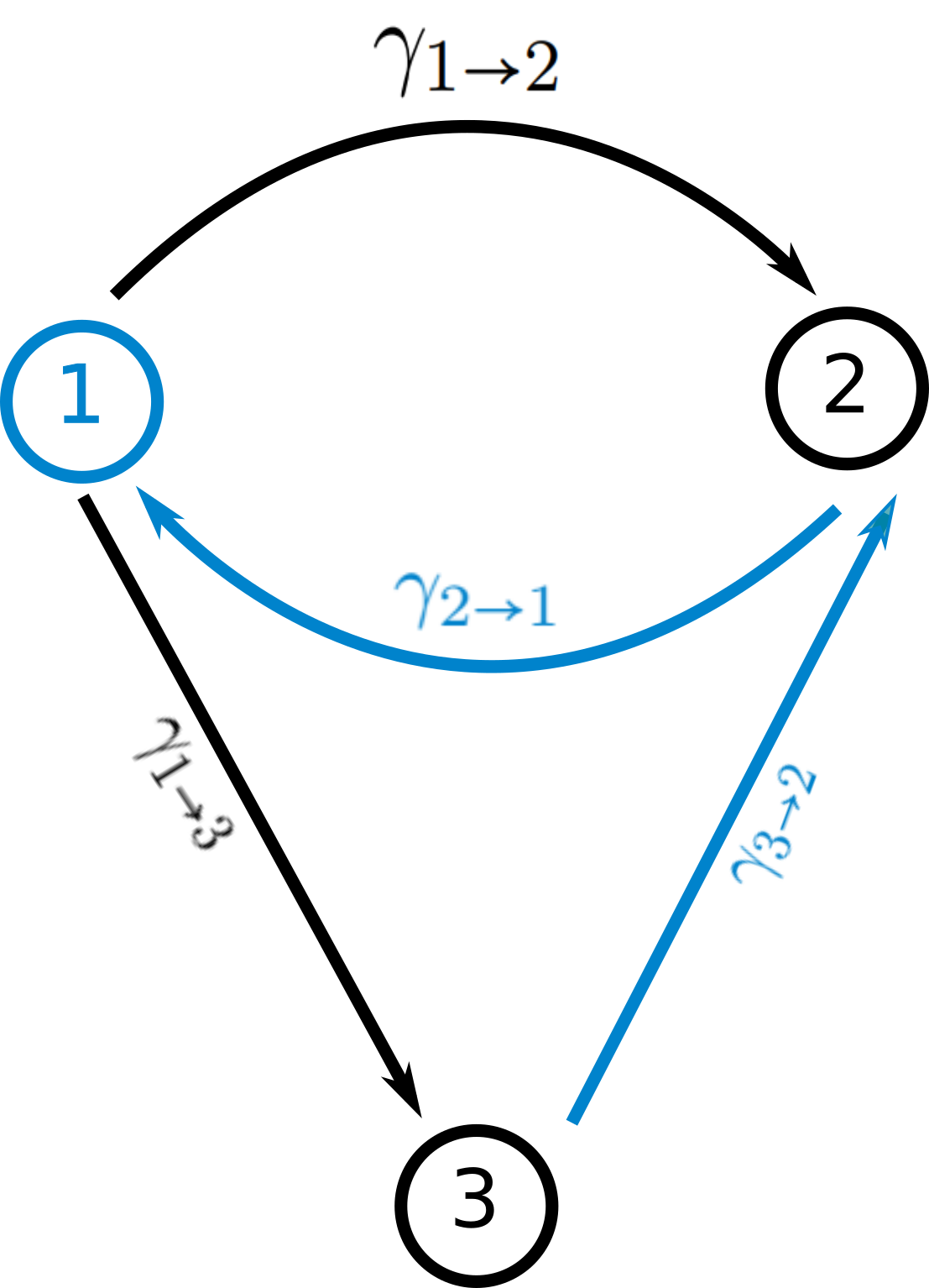} 
\end{subfigure} \hspace*{3mm} \begin{subfigure}{0.25\textwidth}
\subcaption{In-tree rooted in $3$}
\label{Example_StationaryState_InTrees_LeadingTo_3}
 \includegraphics[width=1.0\columnwidth]{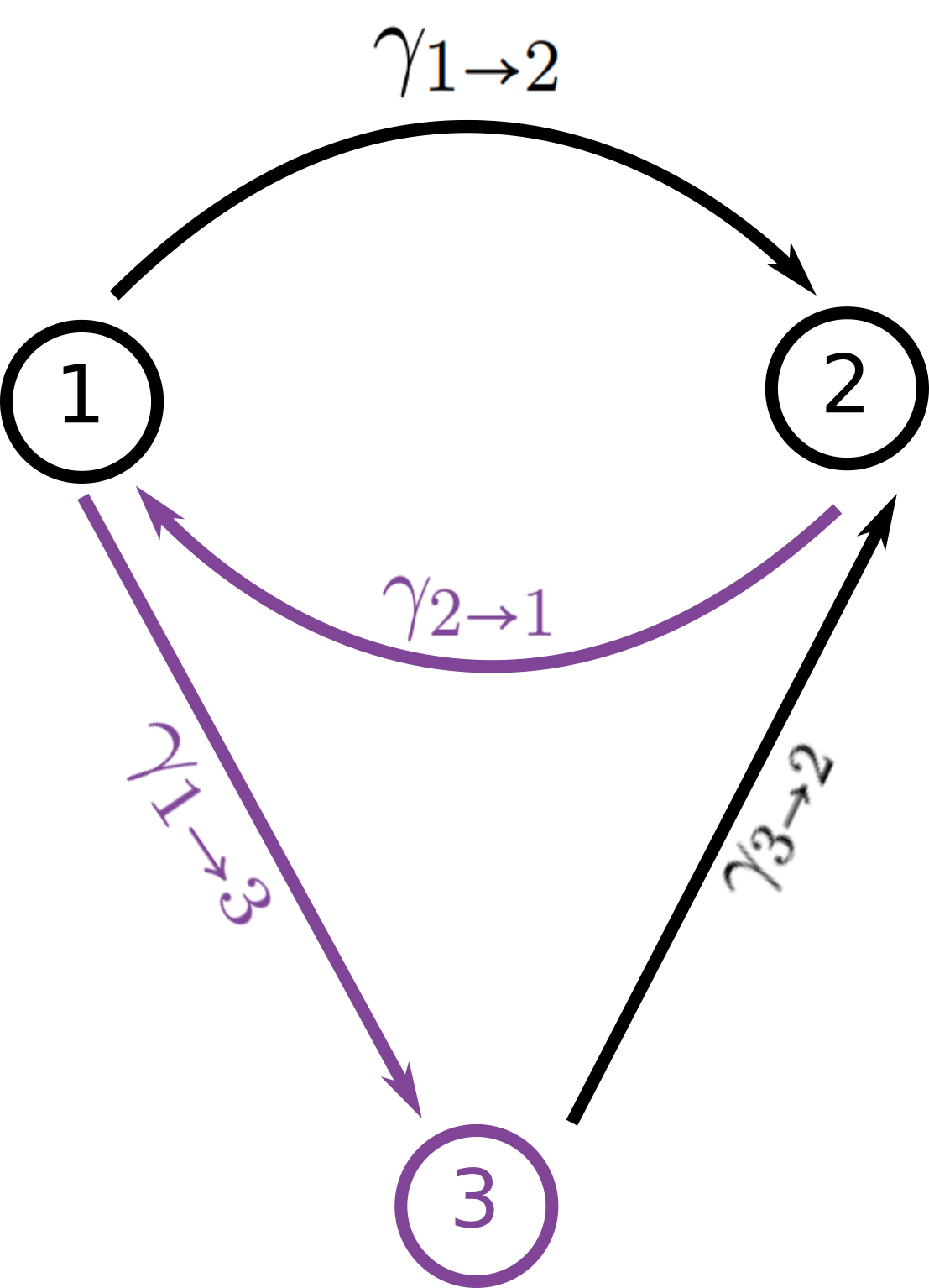} 
\end{subfigure} \\
\vspace*{7mm}
 \begin{subfigure}{0.25\textwidth}
\subcaption{First in-tree rooted in $2$}
\label{Example_StationaryState_InTrees_LeadingTo_2a}
 \includegraphics[width=1.0\columnwidth]{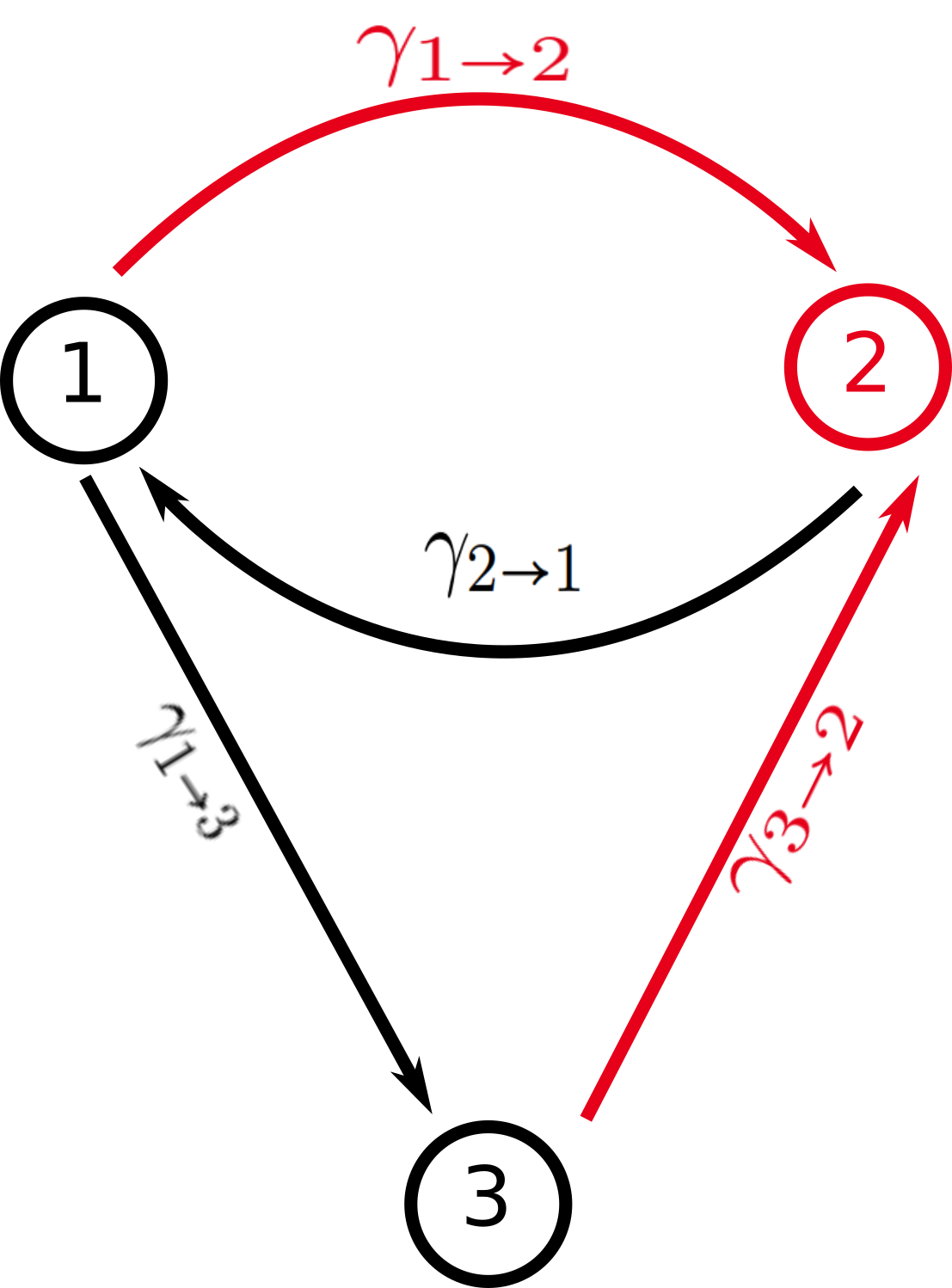}
\end{subfigure} \hspace*{3mm} \begin{subfigure}{0.25\textwidth}
\subcaption{Second in-tree rooted in $2$}
\label{Example_StationaryState_InTrees_LeadingTo_2b}
 \includegraphics[width=1.0\columnwidth]{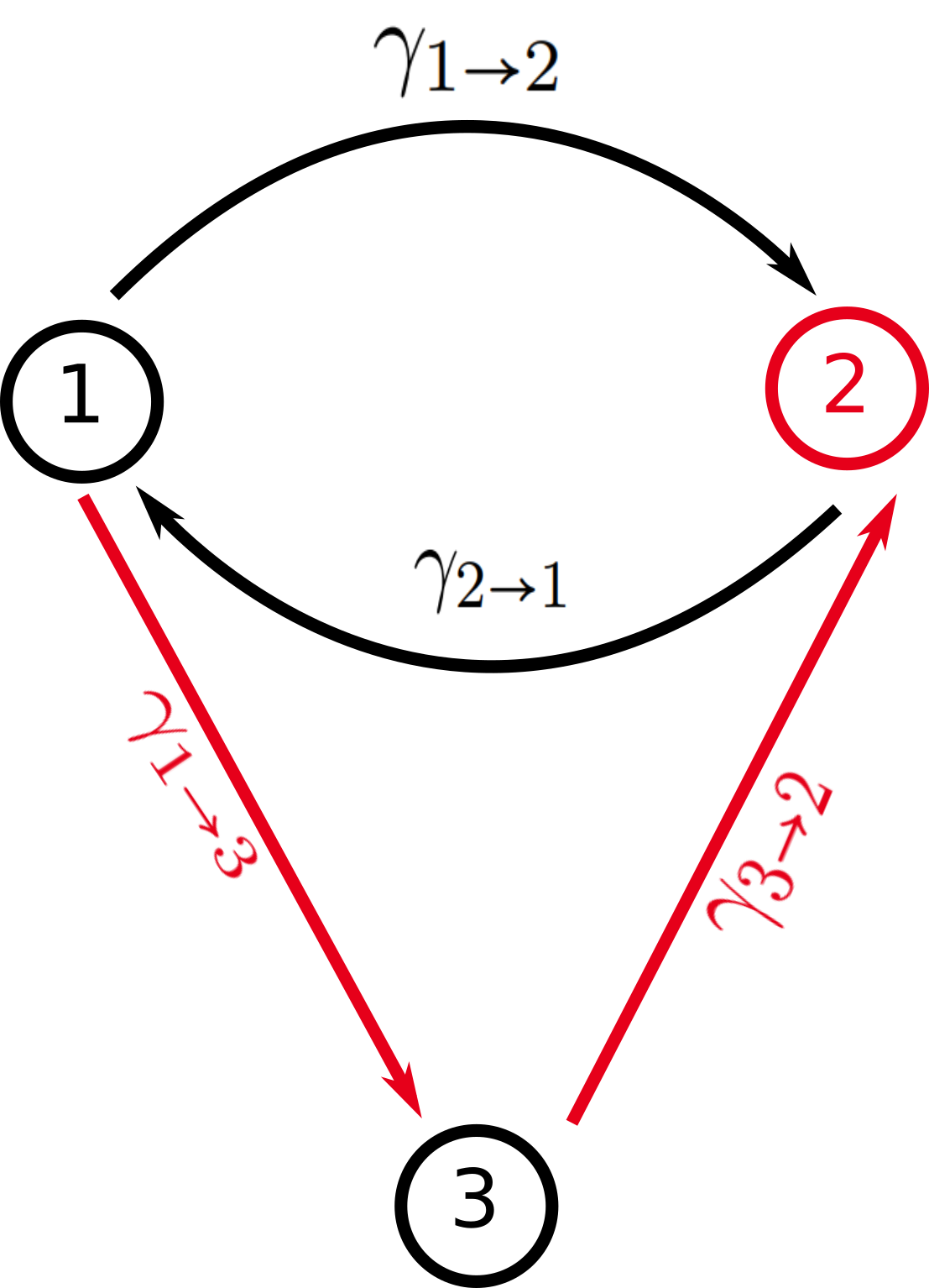} 
 \end{subfigure}
 \end{center}
\end{minipage}
\caption{ Example of a strongly connected network (Figure~\ref{Example_StationaryState_OriginalNetwork}) together with the corresponding in-trees rooted in states number { \color{TUD_2b} $1$ (Figure~\ref{Example_StationaryState_InTrees_LeadingTo_1}) } and { \color{TUD_9b} $3$ (Figure~\ref{Example_StationaryState_InTrees_LeadingTo_3})},  and the two in-trees rooted in state  number {\color{TUD_11a} $2$ (Figure~\ref{Example_StationaryState_InTrees_LeadingTo_2a} and \ref{Example_StationaryState_InTrees_LeadingTo_2b})}. The kernel of $\G$ is the span of a vector whose $i$-th component is the sum over all in-trees rooted in state number $i$ of the product of the rates of all edges that constitute that particular in-tree. In this example we have: \\
$ \Kern(\G) = \Span \left\{ 
\colvec{3}
{ \color{TUD_2b} \g_{3 \to 2} \cdot \g_{2 \to 1} }
{ \color{TUD_9b} \g_{1 \to 2} \cdot \g_{3 \to 2} + \g_{1 \to 3} \cdot \g_{3 \to 2}   }
{\color{TUD_11a} \g_{2 \to 1} \cdot \g_{1 \to 3}} \right\}$. 
}
\label{Example_StationaryState_InTrees}

\end{figure}

For the $k$-th component of that vector you consider an in-tree $T$ spanning over the whole network $\Omega$ rooted in state number $k$ and multiply the rates of all edges $\E(T)$ of $T$. Then you sum over all such in-trees which are rooted in state number $k$. \\
This is done for all component $k \in \{1, \dots, N\}$ that is, for every state number. An illustration of this procedure can be seen in Figure~\ref{Example_StationaryState_InTrees}.

A detailed proof for this theorem can be found in \cite{mirzaev2013laplacian, chaiken1978matrix}. The original statement was first formulated in 1948 by Tutte \cite{tutte1948dissection} . \\

It is possible to find all in-trees in $\Order\left(N + |\E| + |\E|\cdot n\right)$ number of steps \cite{gabow1978finding}, where $N$ is the number of states, $|\E|$ the number of edges and $n$ the number of minimal absorbing sets. 

As a special case, we consider the the microcanonical ensemble of statistical physics \cite{honerkamp2012statistical}
where the rates for a transition and the corresponding reverse transition are identical, $ \g_{i \to j} =  \g_{j \to i}$ for all $i, j \in \{1, \dots, N\}$. From equation \eqref{AnalyticalExpressionForStationaryStatesInMinimalAbsorbingSets} follows that the stationary probability for each state is the same, i.e. $ \bp_\infty = \frac{1}{|\Omega|} (1, \dotsc, 1)^T $: 
For every in-tree rooted at a state $i \in \Omega$, it is possible to construct an in-tree rooted at any other state $j \in \Omega$ by inverting the direction of the transitions. The overall product $\prod\limits_{(i,j) \, \in \,  \E(T))} \g_{i \to j} $ stays the same since the rates for forward and backward transitions are symmetric as is the number of in-trees for each state.

\bibliographystyle{ieeetr}
\bibliography{literature} 

\end{document}